\documentstyle[12pt,axodraw,epsfig]{article}
\begin{document}
\def\gsim{\:\raisebox{-0.75ex}{$\stackrel{\textstyle>}{\sim}$}\:}
\def\lsim{\:\raisebox{-0.75ex}{$\stackrel{\textstyle<}{\sim}$}\:}
\newcommand{\lsp}{\mbox{$\tilde \chi_1^0$}}
\newcommand{\mchi}{\mbox{$m_{\tilde \chi_1^0}$}}
\newcommand{\sig}{\mbox{$\sigma_{\tilde \chi_1^0}$}}
\newcommand{\om}{\mbox{$\Omega_{\tilde \chi_1^0} h^2$}}
\newcommand{\be}{\begin{equation}}
\newcommand{\ben}{\begin{subequations}}
\newcommand{\een}{\end{subequations}}
\newcommand{\beq}{\begin{eqalignno}}
\newcommand{\eeq}{\end{eqalignno}}
\newcommand{\ee}{\end{equation}}
\newcommand{\wt}{\widetilde}
\renewcommand{\thefootnote}{\fnsymbol{footnote}}

\begin{flushright}
TUM--HEP--483/02 \\
October 2002\\
\end{flushright}

\vspace*{2.5cm}
\begin{center}
{\Large \bf  Astroparticle Physics and Colliders}\footnote{Invited plenary
talk at the international workshop on linear $e^+e^-$ colliders (LCWS2002),
Jeju Island, Republic of Korea, August 2002.} \\
\vspace{10mm}
Manuel Drees \\
\vspace{5mm}
{\it Physik Dept., TU M\"unche, D--85748 Garching, Germany}

\end{center}
\vspace{10mm}

\begin{abstract}
In this talk I discuss the interplay between collider physics and four
topics of astro--particle physics: neutrino oscillations, electroweak
baryogenesis, LSP Dark Matter, and ultra--high energy cosmic rays (UHECR).
Some astrophysical scenarios can (only) be tested decisively at colliders.
In other cases input from collider experiments is required to sharpen
predictions for future astro--particle physics experiments, e.g. for
the LSP detection rate or the UHECR spectrum in top--down models.
\end{abstract}
\clearpage

\setcounter{page}{1}
\pagestyle{plain}
\section*{1) Introduction}

The organizers of this conference asked me to give a review talk on
what is commonly known as astro--particle physics. This describes a
vast field of research. I obviously have neither the time nor the
expertise to cover most of it adequately. I therefore decided to
narrow the focus on some topics that have, or at least may have,
fairly direct connections to experiments (to be) performed at high
energy particle colliders. This not only matches the main theme of
this conference, it also suits my personal interests. In the following
four sections I will therefore describe lepton flavor violation (LFV)
as predicted in certain supersymmetric models of neutrino masses that
can explain the observed pattern of neutrino flavor oscillations;
electroweak baryogenesis, which does not work in the Standard Model
(SM) but may work in its supersymmetric extension, the MSSM; the LSP
as cold Dark Matter candidate; and the ``top--down'' interpretation of
ultra--high energy cosmic rays, which posits that these events
originate from the decay or annihilation of ultra--massive particles.

\setcounter{footnote}{0}
\section*{2) Neutrino Oscillations}

It may be appropriate to begin this survey by reminding ourselves that
astro--particle physics experiments, most notably SuperK \cite{1},
succeeded in committing the murder which a great may experimenters
working at colliders had spent twenty--odd years attempting: by
unambiguously showing that muon (anti)neutrinos oscillate into other
flavors, most likely into $\nu_\tau$, the minimal SM was killed, since
it predicts vanishing neutrino masses and hence separately conserved
$e, \, \mu, \, \tau$ lepton numbers. This is a constructive proof that
astro--particle physics experiments can produce results which impact
``mainstream'' particle physics. Similarly, recent SNO data \cite{2}
have shown that $\bar\nu_e$ change into something else somewhere
between the Sun and solar neutrino detectors on Earth, which implies
that electron number is not conserved. This is again most naturally
explained by neutrino flavor oscillations \cite{3}, but in this case
other explanations are still possible \cite{4} (since no ``$L/E''$
plot showing the characteristic energy dependence of neutrino
oscillations has yet been established). We now know that some
neutrinos are massive, and that none of the three lepton flavors is
conserved.

These discoveries are undoubtedly of great importance both conceptually
and for the planning of future Earth--based neutrino oscillation
experiments. However, its relevance to collider physics is a priori
much less clear. In the framework of the SM, small neutrino masses can
most naturally be accommodated by introducing a dimension--5 term in
the Lagrangian, of the form
\be \label{e1}
{\cal L}_\nu = \frac{\lambda_{ij}} {M} \overline{L^C}_i H L_j H + h.c.
\ee
Here $H$ is the Higgs doublet, the $L_i$ are left--handed lepton
doublets ($i,j$ being flavor labels), $M$ is a large mass scale, and
the $\lambda_{ij}$ are numerical coefficients. The data can be
accommodated by allowing $\lambda$ to have non--vanishing
off--diagonal entries, which clearly violates lepton flavor. This LFV
also manifests itself in processes of interest to collider
physicists. For example, the large $\nu_\mu - \nu_\tau$ mixing implied
by atmospheric neutrino oscillations gives rise to $\tau \rightarrow
\mu \gamma$ and $Z \rightarrow \mu^\pm \tau^\mp$ decays through the
one--loop diagrams shown in Fig.~1. However, explicit evaluation of
these diagrams reveals that they are suppressed by powers of the
neutrino mass, leading to predictions for the branching ratios of
these LFV decays many orders of magnitude below the sensitivity of
conceivable experiments.  This result can be generalized: In the SM
augmented by the term (\ref{e1}), or by tiny Yukawa couplings giving
rise to Dirac neutrino masses, LFV at colliders is unobservably small
\cite{6}.

\vspace*{5mm}
\begin{center}
\begin{picture}(400,100)(0,0)
\ArrowLine(75,30)(150,30)
\ArrowLine(225,30)(150,30)
\PhotonArc(150,30)(33,0,180){4}{8.5}
\Photon(100,100)(140,80){4}{5.5}
\Text(65,30)[]{$\tau$}
\Text(235,30)[]{$\mu$}
\Text(130,24)[]{$\nu_\tau$}
\Text(170,24)[]{$\nu_\mu$}
\Text(150,52)[]{$W$}
\Text(90,100)[]{$\gamma,Z$}
\Text(150,31)[]{$\times$}
\end{picture}
\end{center}
\vspace*{-6mm}
\noindent {\bf Fig.1:} {\it Diagram giving rise to $\tau \rightarrow \mu
\gamma$ and $Z \rightarrow \mu \tau$ decays in the SM augmented by the
dimension--5 term (\ref{e1}). The external boson ($\gamma$ or $Z$) has
to be appended in all possible ways, and the $\times$ denotes a
$\nu_\tau \leftrightarrow \nu_\mu$ flavor transition. These diagrams
are suppressed by powers of $m_\nu / M_W$ and are thus unobservably
small.}

\vspace*{8mm}

The situation can be quite different in supersymmetric extensions of
the SM, {\em if} the scale of LFV is below the energy scale where SUSY
breaking is transmitted to the visible sector (i.e., where
``ordinary'' sparticles obtain their masses) \cite{7}. Consider a
supersymmetric see--saw model \cite{5} with three $SU(2) \times
U(1)_Y$ singlet superfields $N_i$ appearing in the superpotential via
\be \label{e2}
{\cal W}_N = f_{ik} H_u L_i N_k + \frac{1}{2} M_{kl} N_k N_l,
\ee
where $H_u$ is the Higgs doublet coupling to up--type quarks. At energies
below the smallest eigenvalue $M_N$ of the matrix $M$, the three
singlets can be integrated out, producing a term of the form (\ref{e1})
in the effective Lagrangian (with $H$ replaced by $H_u$). The coefficients
$\lambda$ of eq.(\ref{e1}) are then bilinear functions of the new
Yukawa couplings $f$ appearing in eq.(\ref{e2}), leading to neutrino
masses of order
\be \label{e3}
m_\nu \sim |f|^2 \langle H_u^0 \rangle^2 / M_N .
\ee

The crucial difference between the SM and the MSSM is that the latter
contains additional parameters (besides the $\lambda_{ij}$) that can
violate lepton flavor: the soft breaking masses and trilinear $A$
parameters in the slepton sector. High--scale LFV therefore does {\em
not} necessarily decouple in the MSSM. In particular, even if we
assume that slepton masses are universal at some high scale $M_U >
M_N$, with value $\overline{m}_{\tilde L}$, the LFV violation encoded
in the $f_{ij}$ will radiatively induce LFV slepton masses of order
\cite{8}
\be \label{e4}
\delta \left(m^2_{\tilde L}\right)_{ij} \sim \frac{1} {8 \pi^2}
\sum_k f^*_{ki} f_{kj} \left( 2 \overline{m}_{\tilde L}^2 + m^2_{H_u} +
|A_{\tilde L}|^2 \right) \log \frac {M_U}{M_N};
\ee
the exact (1--loop) RGE can be found in \cite{9}. Eqs.(\ref{e3}) and
(\ref{e4}) together imply that this new source of LFV scales like
$m_\nu M_N / M_W^2 \cdot \log(M_U/M_N)$. For given neutrino mass this
becomes maximal for $M_N$ only slightly below $M_U$. In other words, this
new physics effect {\em increases} with increasing energy scale where
the effect originates! 

These flavor off--diagonal slepton masses can contribute to $Z
\rightarrow \mu^\pm \tau^\mp$ and $\tau \rightarrow \mu \gamma$ decays
through the one--loop diagrams shown in Fig.~2 \cite{7}. There is
again some GIM--type cancellation; however, the relevant suppression
factor is now $\left( m^2_{\tilde L} \right)_{\mu\tau} /
\overline{m}^2_{\tilde L}$, which is usually much bigger than $m_\nu/M_W$.
This suppression is nevertheless crucial for bringing supersymmetric
see--saw models with large neutrino mixing angles and high--scale
SUSY breaking into agreement with stringent experimental limits on
$\tau \rightarrow \mu \gamma$ and $\mu \rightarrow e \gamma$ decays.

\vspace*{5mm}
\begin{center}
\begin{picture}(800,100)(0,0)
\ArrowLine(50,30)(92,30)
\ArrowLine(200,30)(158,30)
\DashLine(92,30)(158,30){2}
\CArc(125,30)(33,0,180)
\Photon(75,100)(115,80){4}{5.5}
\Text(40,30)[]{$\tau$}
\Text(210,30)[]{$\mu$}
\Text(105,24)[]{$\tilde\tau_L$}
\Text(145,24)[]{$\tilde\mu_L$}
\Text(125,52)[]{$\tilde \chi^0_i$}
\Text(65,100)[]{$\gamma,Z$}
\Text(125,31)[]{$\times$}
\Text(250,50)[]{\large\bf{+}}
\ArrowLine(300,30)(342,30)
\ArrowLine(450,30)(408,30)
\DashLine(342,30)(408,30){2}
\CArc(375,30)(33,0,180)
\Photon(325,100)(365,80){4}{5.5}
\Text(290,30)[]{$\tau$}
\Text(460,30)[]{$\mu$}
\Text(355,24)[]{$\tilde\nu_\tau$}
\Text(395,24)[]{$\tilde\nu_\mu$}
\Text(375,52)[]{$\tilde \chi^-_i$}
\Text(315,100)[]{$\gamma,Z$}
\Text(375,31)[]{$\times$}
\end{picture}
\end{center}
\vspace*{-6mm}
\noindent {\bf Fig.2:} {\it SUSY contributions to $\tau \rightarrow
\mu \gamma$ and $Z \rightarrow \mu \tau$, where $\times$ now denotes a
slepton flavor transition. These contributions are only suppressed by
powers of $\left( \delta m_{\tilde L} \right)_{\mu\tau} /
\overline{m}_{\tilde L}$.}

\vspace*{8mm}

This suppression of loop--induced LFV decays may also offer the chance
to observe {\em tree--level} LFV effects in sparticle production and
decay at high energy colliders. For example, one can search for
sneutrino pair production at lepton colliders, followed by $\tilde
\nu_i \rightarrow \ell_k \tilde \chi^-_1 \rightarrow \ell_k j j \tilde
\chi_1^0$ decays. This can give rise to final states containing
$\mu^\pm \tau^\mp$ pairs together with up to four jets and missing
energy/momentum, which have very little background \cite{8}. These LFV
signals can be understood as being to due $\tilde L$ flavor
oscillations \cite{10}. They are therefore only suppressed by powers
of $\delta m_{\tilde L} / \overline{\Gamma}_{\tilde L}$, where
$\overline{\Gamma}_{\tilde L} \sim \alpha \overline{m}_{\tilde L}$ is
the average decay width of the oscillating sleptons. Since the largest
new Yukawa couplings $f$ are assumed to (again) occur in the third
generation, and $\nu_\tau$ is known to strongly mix with $\nu_\mu$
but only weakly with $\nu_e$, for given luminosity $\mu^+\mu^-$ colliders
offer better sensitivity for these $\tilde \nu$ oscillation signals than
$e^+e^-$ colliders do \cite{8}; however, it now seems likely that a
muon collider, if it can be built at all, will have far smaller
luminosity than that currently foreseen for linear $e^+e^-$ colliders.
Slepton oscillations can also be tested in slepton production at
hadron colliders \cite{10,11}, as well as through the production of
heavier neutralinos and charginos followed by their leptonic decay
\cite{10,11,12}. These signals can probe regions of parameter space that
are compatible with bounds on LFV $Z, \, \tau$ and $\mu$ decays. 
Moreover, analysis of such signals allows (at least in principle) to
determine the quantities $f_{ij}$ and $M_{ij}$ separately, rather than
only the combinations of these parameters that produces the effective
neutrino mass term (\ref{e1}) \cite{12a}, {\em if} the slepton masses
at scale $M_U$ are known.

However, all these SUSY--mediated LFV effect vanish if the scale of
LFV is higher than the scale where SUSY breaking is mediated to the
visible sector, as is the case e.g. in most see--saw models with
gauge--mediated supersymmetry breaking. On the other hand,
supersymmetry also permits an entirely novel way to generate neutrino
masses, if $R-$parity is broken. In general one neutrino will become
massive through mixing with neutralinos, while the other neutrino
masses are generated through loop diagrams. Many models of this kind
predict relations between neutrino oscillations and decays of the
lightest superparticle (LSP), where the latter may include LFV modes;
this again offers the possibility to detect LFV at colliders.  I will
not discuss this approach here, since it has been covered at this
conference by E.J. Chun \cite{13}.

\setcounter{footnote}{0}
\section*{3) Electroweak Baryogenesis}

The Universe around us contains baryons and leptons, but very few
antibaryons and antileptons.\footnote{The small $e^+$ and $\bar{p}$
components in the primary cosmic ray flux have most likely been
produced in the collisions of relativistic $e^-$ and $p$ with ordinary
matter.} Analyses of Big Bang nucleosynthesis \cite{14} show that the
baryon to photon ratio,
\be \label{e5}
\eta \equiv \frac {n_b - n_{\bar b}} {n_\gamma},
\ee
must be a few times $10^{-10}$. In other words, there is a very small,
but non--vanishing, excess of particles over antiparticles in the
Universe. This is not an existence proof of ``new physics'', since
such a small asymmetry could simply be imposed as an initial condition
on the Universe. However, a {\em dynamical} explanation of such a
small excess, starting from a Universe with vanishing total baryon and
lepton number, would certainly be more attractive.

In his seminal 1967 paper \cite{15}, Sakharov listed three necessary
conditions any dynamical mechanism must satisfy: it must (obviously)
violate baryon number; it must violate both C and CP, since the baryon
number generator $B$ is odd under both C and CP, so that $\langle 0 |
B| 0 \rangle$ would have to vanish if $C$ or $CP$ were good quantum
numbers; and it must be active out of thermal equilibrium, since
otherwise the rate for any reaction $A \rightarrow B$ equals that of
the inverse reaction $B \rightarrow A$, making it impossible to create
a net baryon number. 

Somewhat later it was realized \cite{16} that a fourth condition has
to be satisfied. The reason is that the SM predicts electroweak baryon
number violating ``sphaleron'' transitions. Roughly speaking, these
correspond to transitions between degenerate (ground) states with
different ``topology'' [in the space of $SU(2)$ gauge
configurations]. These states are separated by ``potential barriers''
whose height is given by the vev that breaks the electroweak gauge
symmetry. At zero temperature these transitions are suppressed by a
tunneling factor $\exp\left( - 16 \pi^2/ g_2^2 \right)$, where $g_2$
is the $SU(2)$ gauge coupling, and can thus safely be
neglected. However, these transitions are in thermal equilibrium at
temperatures $T \geq 100$ GeV. Note that these transitions violate
both baryon and lepton number, but conserve $B-L$. The fourth
condition that any dynamical explanation of the baryon asymmetry of
the Universe (BAU) must satisfy in order to be ``immune'' to sphaleron
transitions is therefore the following: it must either produce a
non--vanishing $B-L$, or else produce non--vanishing $B$ at
temperature (well) below 100 GeV.

Of course, the SM also violates C and CP. Together with the
sphaleron--induced violation of baryon number, this satisfies two of
the three Sakharov conditions. Moreover, the third condition,
deviation from thermal equilibrium, is satisfied during phase
transitions.\footnote{For example, boiling water is a non--equilibrium
process. This can be seen from the fact that boiling water behaves
quite differently from condensing steam, even if temperature and
pressure are the same.} Such a phase transition might be associated
with electroweak symmetry breaking. To see this, consider the
temperature dependent effective potential of the Higgs
field\footnote{$\phi$ only describes the field acquiring a vev,
e.g. the real part of the neutral component of the Higgs doublet in
the SM, or a linear combination of the two analogous real parts in the
MSSM.} $\phi$:
\be \label{e6}
V_{\rm eff}(\phi, T) \simeq \frac {\gamma T^2 - m^2}{2} \phi^2 - E T \phi^3
+ \frac{\lambda}{2} \phi^4.
\ee
Here $-m^2$ (the infamous negative mass--squared) and $\lambda$
describe the potential at $T=0$, while $\gamma$ and $E$ describe the
leading effects of the thermal plasma. The qualitative behavior of
this potential is illustrated in fig.~3. Since $\gamma$ and $E$ are
both positive, at sufficiently high temperature the origin ($\phi=0$)
will be a minimum, not a maximum. Indeed, for $T > T_c$ it will be the
absolute minimum, i.e. the $SU(2) \times U(1)_Y$ symmetry will not be
broken. However, a second minimum may exist at $\phi = \phi_0 \neq 0$,
which becomes more prominent as $T$ decreases. At some critical
temperature $T_c$ this second minimum satisfies $V( \phi_0 , T) = 0$,
i.e. it is degenerate with the minimum at the origin. It is easy to
show that this requires
\be \label{e7}
\phi_c \equiv \phi_0 (T = T_c) = \frac {E T_c} {\lambda}.
\ee
At slightly lower temperature the Universe begins to tunnel from $\phi
= 0$ to $\phi = \phi_0$. The rate of (thermal) sphaleron transitions
at this tunneling temperature depends very sensitively on $\phi_0$ as
well as $T$. It can be shown that these transitions are sufficiently
suppressed if
\be \label{e8}
\phi_c \gsim T_c,
\ee
which requires $E \gsim \lambda$ from eq.(\ref{e7}). Recall that
$\lambda$ is proportional to the square of the mass of the (lightest
CP--even) Higgs boson. The experimental lower bound on the mass of
(SM--like) Higgs bosons therefore translates into a {\em lower} bound
on $\lambda$. Moreover, $E$ receives significant contributions only
from bosons with mass not greatly in excess of $T$. In the SM, which
contains few elementary bosons, condition (\ref{e8}) is therefore
badly violated for all experimentally allowed values of $\lambda$;
indeed, here $E/\lambda$ is so small that no proper phase transition
occurs \cite{16a}.

\setcounter{figure}{2}
\begin{figure}[htb]
\vspace*{6mm}
\hspace*{15mm}
\mbox{
\epsfig{file=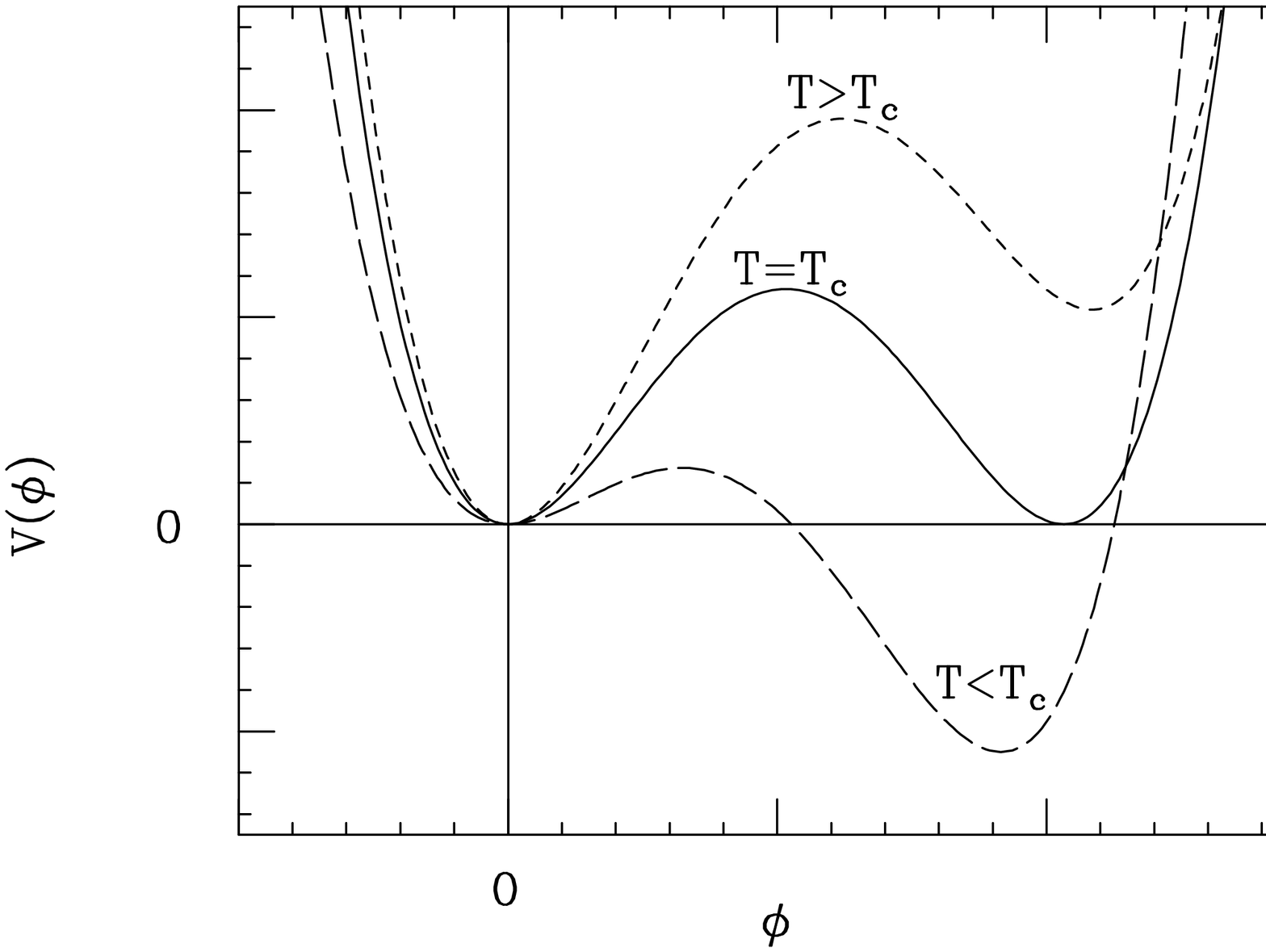,width=0.7\textwidth} }
\vspace*{-2mm}
\end{figure}
\noindent {\bf Fig.3:} {\it The temperature--dependent effective
potential at temperatures above, at, and below the critical
temperature $T_c$.}

\vspace*{5mm}

The situation is somewhat more favorable in the MSSM. The experimental
lower bound on $\lambda$ is (slightly) weaker than in the SM. More
importantly, there are many more elementary scalars in the theory. The
potentially most important ones are the $\tilde t$ squarks, due to
their large (Yukawa) couplings to $\phi$. Recall, however, that
$\tilde t$ can only contribute significantly to $E$ if $m_{\tilde t}$
does not greatly exceed $T_c$. At the same time, some $\tilde t$
squarks must be quite heavy to produce sufficiently large loop
corrections to the mass of the lightest CP--even Higgs boson, raising
it from the tree--level upper bound of $M_Z$ to at least the
experimental lower bound. Moreover, electroweak precision data imply
lower bounds of several hundred GeV on the masses of $SU(2)$ doublet
$\tilde t$ and $\tilde b$ squarks. This leads one to consider
scenarios with heavy $SU(2)$ doublet squarks, but light $SU(2)$
singlet $\tilde t_R$. Indeed, it can be shown \cite{17} that some
region of parameter space satisfying the constraint (\ref{e8}) still
exists, with $m_{\tilde t_L} \gsim 1$ TeV, $m_{\tilde t_R} \leq m_t$
and $m_h \leq 117$ GeV; the latter two predictions can easily be
tested at future collider experiments.

The existence of a light $\tilde t_R$ and a fairly light Higgs boson
ensures that condition (\ref{e8}) is satisfied, which in turn implies
that the third Sakharov condition is met. As well known, the second
Sakharov condition is satisfied already in the SM, where CP violation
originates in complex Yukawa couplings, and manifests itself in the
complex phase of the KM matrix. However, it turns out that this source
of CP violation is too weak to produce a sufficiently large baryon
asymmetry in the present framework. One needs large \cite{18}, perhaps
even maximal \cite{19}, CP violation in the chargino mass matrix,
together with relatively light charginos. This prediction can also be
tested through future precision measurements, e.g. at $e^+e^-$ linear
colliders \cite{20}.

Electroweak baryogenesis can thus easily be falsified by future
collider experiments. This remains true in extensions of the MSSM
which contain $SU(2) \times U(1)_Y$ singlet Higgs superfields
\cite{21}: although this makes it somewhat easier to satisfy the
constraint (\ref{e8}), the required CP violation in the chargino
system is essentially the same as in the MSSM. Unfortunately, other
proposed explanations of the BAU cannot be tested in the
laboratory. For example, in ``leptogenesis'' \cite{22} one creates a
lepton asymmetry at high $T$, which is partially converted to a baryon
asymmetry by electroweak sphaleron transitions. The lepton asymmetry
is generated in the production and decay of ``right--handed'' [$SU(2)
\times U(1)_Y$ singlet] (s)neutrinos with mass typically exceeding
$10^8$ GeV, well beyond the reach of conceivable colliders. The same
heavy neutrinos influence the mass matrix of the light neutrinos
through the see--saw mechanism \cite{5}; however, the CP violating
phase required for leptogenesis is not necessarily related to the
phase that can (in principle) be measured in neutrino oscillation
experiments \cite{23} (but in supersymmetric models with universal
soft breaking terms at some high scale, it can be measured in
low--scale LFV processes \cite{23a}). Similarly, the Affleck--Dine
mechanism \cite{24} produces a baryon or lepton asymmetry from a
sfermion condensate just after the end of inflation; its dynamics is
governed by parameters that cannot be measured at colliders. However,
all proposed dynamical explanations of the BAU require the existence
of new CP violating phases. We just saw that in many cases these
phases cannot be measured at collider energies; nevertheless the
existence of new sources of CP violation can serve as motivation to
search for CP violation wherever experimentally feasible, including
channels where no CP violation is expected in the SM.

\setcounter{footnote}{0}
\section*{4) Neutralino Dark Matter}

The existence of Dark Matter (DM) in the Universe is quite well
established. Historically the first evidence came from the analysis of
the motion of gravitationally bound systems. The simplest example of
this kind is given by the well--known galactic rotation curves. An
object on a circular orbit (with radius $R$) around a spherical galaxy
must have orbital velocity $v(R) = \sqrt{G_N M(R) / R}$, where $G_N$
is Newton's constant and $M(R)$ is the mass inside the orbit. If a
galaxy's mass was confined to its luminous parts, one should thus find
$v(r) \propto 1/\sqrt{R}$ for sufficiently large $R$; however,
observation yields essentially flat rotation curves, $v(R) \sim
$constant, which implies that the luminous part of the galaxy is
embedded in a much larger dark halo. Analogous arguments can be made
for the motion of galaxies inside clusters. More recently,
measurements of the anisotropy of the cosmic microwave background on
degree scales as well as observations of type--Ia supernovae at high
redshift have determined the total matter density of the Universe to
fall in the range \cite{25}
\be \label{e9}
0.1 \lsim \Omega_{\rm matter} h^2 \lsim 0.3.
\ee
Here, $\Omega_X$ denotes the mass (or energy) density of component $X$
in units of the critical density (i.e. $\sum_X \Omega_X = 1$
corresponds to a flat Universe), and $h$ is the Hubble constant in
units of 100 km/(sec$\cdot$Mpc). $\Omega_{\rm matter}$ includes all
forms of non--relativistic matter, but analyses of Big Bang
nucleosynthesis have shown that $\Omega_{\rm baryon} h^2 < 0.05$
\cite{25}, so that the range (\ref{e9}) essentially also holds for the
DM component.

The lightest neutralino \lsp\ is probably the best motivated, and
certainly the most widely studied \cite{26}, particle physics DM
candidate. In many SUSY models \lsp\ is the lightest superparticle
(LSP) over much of parameter space; it is then absolutely stable, if
$R-$parity is conserved. It is also neutral, which is required
experimentally for any massive particle that contributes significantly
to DM; charged particles would bind to nuclei, thereby forming
isotopes with exotic mass--to--charge ratios, which have not been
found \cite{27}. Finally, the present thermal LSP relic density is at
least within a few orders of magnitude of the desired range
(\ref{e9}).

Let us investigate this last argument in some more detail, since it is
often used to single out regions of SUSY parameter space as being
``cosmologically favored''. The basic assumption is that \lsp\ was in
thermal equilibrium with the hot plasma of SM particles after the last
period of entropy production, which in simple cosmological models
occurred at the end of inflation. This requirement is satisfied if the
``re--heat'' temperature at the end of this epoch of entropy
production exceeded $0.1 \mchi$; note that the scale of inflation is
supposed to be around $10^{13}$ GeV, so re--heat temperatures in
excess of $0.1 \mchi$ are easy to achieve. Neutralinos will then stay
in equilibrium as long as the rate of reactions that change the number
of LSPs is larger than the Hubble expansion rate. $R-$parity
conservation implies that LSPs can only be created or annihilated in
pairs, so that the equilibrium condition reads
\be \label{e10}
n_{\tilde \chi_1^0}(T) \langle v \sig \rangle > H(T) .
\ee
Here, $n_{\tilde \chi_1^0}$ is the LSP number density, \sig\ the total
\lsp\ pair annihilation cross section, $v$ the relative velocity
between the two LSPs in the cm frame, and $\langle \dots \rangle$
denotes thermal averaging.\footnote{Note that LSPs will stay in
kinetic equilibrium with the SM plasma, i.e. will have a thermal
velocity distribution, long after they drop out of chemical
equilibrium.} At temperatures $< \mchi$, $n_{\tilde \chi_1^0} \sim
\left( \mchi T \right)^{3/2} \exp \left( - \mchi/T \right)$ is
exponentially suppressed, whereas the Hubble parameter in the
radiation--dominated era, $H(T) \sim T^2/M_{\rm Pl}$, $M_{\rm Pl}
\simeq 2.4 \cdot 10^{18}$ GeV being the Planck mass, only decreases
like a power of $T$. Clearly the condition (\ref{e10}) can therefore
not be satisfied at arbitrarily low temperatures. Due to the $1/M_{\rm
Pl}$ factor in $H(T)$, LSPs stay in chemical equilibrium with the SM
plasma until they are quite non--relativistic; freeze--out
(decoupling) typically occurs at $T_F \simeq \mchi/20$. Clearly
decoupling will occur later, at a smaller value of $n_{\tilde
\chi_1^0}$, for larger values of $\langle v \sig \rangle$. Indeed, in
this scenario one finds that the present \lsp\ relic mass density is
essentially inversely proportional to $\langle v \sig \rangle$; this
is quite intuitive: the stronger LSPs annihilate, the fewer are left
over. Moreover, the relic density has the right order of magnitude for
roughly weak--scale cross sections, $\sig \sim 0.01$ pb or so. This at
the very least establishes an intriguing coincidence between particle
physics and cosmology.

Dimensional analysis shows that \sig\ should decrease, i.e. \om\
should increase, like the square of the sparticle mass
scale. Moreover, two identical Majorana fermions can annihilate into a
massless fermion--antifermion pair only from a $P-$wave (or higher)
initial state.\footnote{This statement is strictly true only for a
CP--invariant theory. However, bounds on the electric dipole moments
of the neutron and electron tightly constrain the CP--violating phases
relevant to \lsp\ annihilation.} Since \lsp s are already quite
non--relativistic at freeze--out, this reduces the thermal average of
$v \sig$ by about an order of magnitude. If the LSP is mostly a bino,
which is true over most of parameter space in many SUSY models, the
upper bound in (\ref{e9}) can only be satisfied in three different
regions of parameter space \cite{28,29}:

\begin{itemize}

\item {\it The bulk region:} If $\lsp \simeq \tilde b$, the biggest
contribution to \sig\ comes from $\ell^+\ell^-$ final states, which in
turn receive the most important contributions from $\tilde \ell_R$ (or
$\tilde \tau_1$) exchange in the $t-$ or $u-$channel ($\ell = e, \,
\mu, \, \tau$). One reason for this is that $\tilde \ell_R$ has the
largest hypercharge, and hence the strongest coupling to $\tilde b$,
of all sfermions. Moreover, since $\tilde \ell_R$ is an $SU(3) \times
SU(2)$ singlet, its mass only receives rather small contributions from
gaugino loop diagrams; such diagrams always increase the sfermion
mass, and are e.g. responsible for the fact that squarks without
sizable Yukawa couplings cannot be much lighter than gluinos. Here
$\om \leq 0.5$ requires \cite{29a} $\mchi, \, m_{\tilde \ell_R} \leq
250$ GeV, which can readily be tested at a first--generation linear
collider.

\item {\it The co--annihilation region:} The $\tilde \ell_R$ (or
$\tilde \tau_1$) exchange contribution to \sig\ is maximal if
$m_{\tilde \ell_R}$ (or $m_{\tilde \tau_1}$) is essentially equal to
\mchi. In such a situation \om\ can no longer be estimated from \lsp\
pair annihilation alone \cite{30}. If another superparticle $\tilde P$
is close in mass to \lsp, reactions of the kind $ \lsp + f
\leftrightarrow \tilde{P} + f'$, with rate $\propto \exp \left( -
m_{\tilde P} / T \right)$, will stay in equilibrium much longer than
reactions of the kind $\lsp + \lsp \leftrightarrow f + \bar f$, which
have rate $\propto \exp \left( - 2 \mchi / T \right)$; here $f, f'$
are some SM particles with mass $\lsim T$. In this case the total
number of superparticles, which determines the final LSP relic
density, can also be reduced by $\lsp \tilde P$ co--annihilation or
$\tilde P \bar{\tilde P}$ annihilation. Note that the $\tilde P$
number density is suppressed by the Boltzmann factor $\exp \left[ -
\left( m_{\tilde P} - \mchi \right) / T \right]$ relative to the \lsp\
number density. Since the relevant temperature is $T \simeq T_F \lsim
\mchi/20$, this Boltzmann suppression implies that co--annihilation
can only be important for scaled mass difference $\left( m_{\tilde P}
- \mchi \right) / \mchi \lsim 0.2$ or so. On the other hand, the total
co--annihilation cross section can be significantly larger than the
\lsp\ pair annihilation cross section. In case of co--annihilation
with $\tilde \ell_R$ (or $\tilde \tau_1$), the enhancement is
primarily due to the fact that co--annihilation, as well as
annihilation of slepton pairs, can occur from $S-$wave initial states;
it can thus reduce \om\ by about one order of magnitude, for vanishing
mass difference \cite{31}. On the other hand, co--annihilation with a
light $\tilde t_1$ squark, which has strong QCD and Yukawa
interactions, can reduce the relic density by more than three orders
of magnitude \cite{32}. Co--annihilation with sleptons can therefore
``only'' increase the upper bound on slepton masses by about a factor
3.5; this range may still be covered by future $e^+e^-$ colliders. On
the other hand, co--annihilation with $\tilde t_1$ can bring a
bino--like LSP as heavy as 3 TeV into agreement with the constraint
(\ref{e9}), i.e. it allows to construct cosmologically allowed SUSY
spectra well beyond the reach of any currently planned collider;
recall that in models with gaugino mass unification a bino mass of 3
TeV implies a gluino mass near 20 TeV!

\item {\it The Higgs pole region:} The \lsp\lsp\ annihilation cross
section can be greatly enhanced if some $s-$channel diagram becomes
resonant. The potentially most important such diagrams involve the
exchange of the neutral CP--odd Higgs boson $A$, since it can couple
to an $S-$wave initial state. If $2 \mchi \sim m_A$ the annihilation
cross section scales like $1/\Gamma_A^2$ rather than $1/m^2_{\tilde
\chi_1^0}$, which introduces an enhancement factor $1/\alpha^2 \sim
10^4$. This over--compensates the small coupling (of order $g_Y \sin
\theta_W M_Z / |\mu|$) of bino--like neutralinos to Higgs bosons, and
thus again allows to construct cosmologically acceptable scenarios
with $\mchi > 1$ TeV.

\end{itemize}

This discussion shows that \om\ depends on many details of the SUSY
spectrum, i.e. on many soft breaking parameters. It is thus usually
analyzed in models that describe the entire spectrum in terms of a
small number of free parameters, the most common one being the minimal
SUGRA model (sometimes called cMSSM). Here one postulates a common
scalar mass $m_0$, a common gaugino mass $m_{1/2}$, and a common
trilinear scalar interaction parameter $A_0$, at the high ``input''
scale $M_{GUT} \simeq 2 \cdot 10^{16}$ GeV. The ratio of vevs
$\tan\beta$ is also a free parameter, as is the sign of $\mu$, while
$|\mu|$ is fixed by the requirement $M_Z = 91.2$ GeV. In this scenario
the bulk region is limited by $m_0 \lsim 250$ GeV, $m_{1/2} \lsim 350$
GeV, with little dependence\footnote{Unless $\tan\beta$ is very large,
in which case $L-R$ mixing in the $\tilde \tau$ sector reduces \om\
significantly; this mixing depends on both $\tan\beta$ and $A_0$ (in
the latter case, mostly via $|\mu|$).} on $A_0$ and $\tan\beta$. $\lsp
- \tilde t_1$ co--annihilation is difficult to achieve in this model
\cite{29,33}, but co--annihilation with sleptons does occur over a
narrow strip of gaugino masses satisfying $m_{1/2} \simeq 2.5 m_0$,
the exact location depending on $A_0$ and $\tan\beta$. In this model
the Higgs pole region only exists \cite{34} at large $\tan\beta \gsim
50$. This model allows a fourth region where the constraint (\ref{e9})
can be satisfied:

\begin{itemize}
\item {\it The focus point region:} If $m_0^2 \gg m_{1/2}^2$, the soft
breaking contribution to the mass of the Higgs boson that couples to
the top quark is very small at the weak scale, independent of its
value at the input scale (this is the ``focusing'') \cite{35}. In
this case $|\mu|$, the supersymmetric contribution to the Higgs boson
masses, also comes out to be small, so that \lsp\ has significant,
perhaps even dominant, higgsino components. This implies sizable
annihilation cross sections into final states containing electroweak
gauge and/or Higgs bosons. In the extreme case of higgsino--like LSP
this again allows $\mchi > 1$ TeV \cite{35a}; note that annihilation
now occurs from an $S-$wave initial state, via $SU(2)$ [rather than
$U(1)_Y$] couplings.

\end{itemize}

The situation is illustrated in Fig.~4, which shows (in green) the
region in the $(m_0, m_{1/2})$ plane satisfying $\om \leq 0.5$ for
$A_0 = 0, \, \mu > 0$ and $\tan\beta = 10$ (left) and 60 (right). In
the left panel, the ``bulk'' region is near the origin (and thus
severely constrained by collider experiments), while the
``co--annihilation'' and ``focus point'' regions are the nearly
horizontal and nearly vertical strip, respectively. These regions also
exist at $\tan\beta = 60$ (right), but are supplemented by the much
larger ``Higgs pole'' region. This last region is rather wide, since
the thermal motion allows resonant $A-$exchange even if \mchi\ is well
below $m_A/2$. These plots, which are updates of ref.\cite{29},
vividly illustrate that even in this very constrained model the
condition (\ref{e9}) does not allow to derive useful upper bounds on
sparticle masses, although it does exclude large regions of parameter
space. These figures also show experimentally excluded regions (grey),
the region where 113 GeV $\leq m_h \leq$ 117 GeV in accordance with
the 2$\sigma$ LEP evidence for an SM--like neutral Higgs boson (red),
and the region compatible with $g_\mu - 2$ at the $2\sigma$ level
(blue). These laboratory ``discoveries'', if taken seriously, do
impose upper limits on sparticle masses; in particular, a TeV--scale
$e^+e^-$ collider would have to discover $\lsp \tilde \chi_2^0$
production in the overlap region (black) where all constraints are
satisfied.

\vspace*{5mm} \noindent $m_0$

\begin{figure}[htb]
\vspace*{-11mm} \hspace*{5mm} 
\mbox{
\epsfig{file=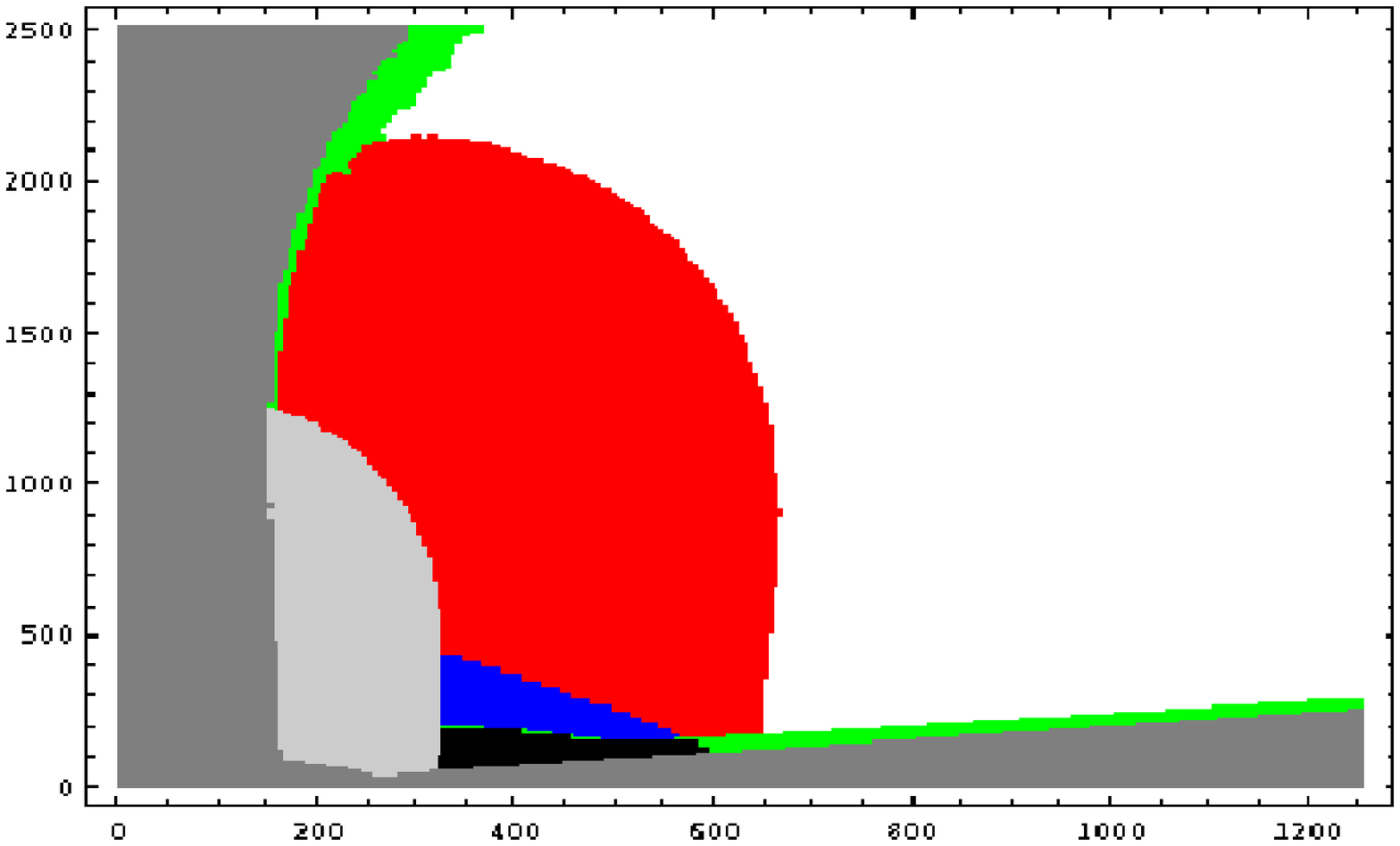,width=0.45\textwidth} 
\hspace*{5mm}
\epsfig{file=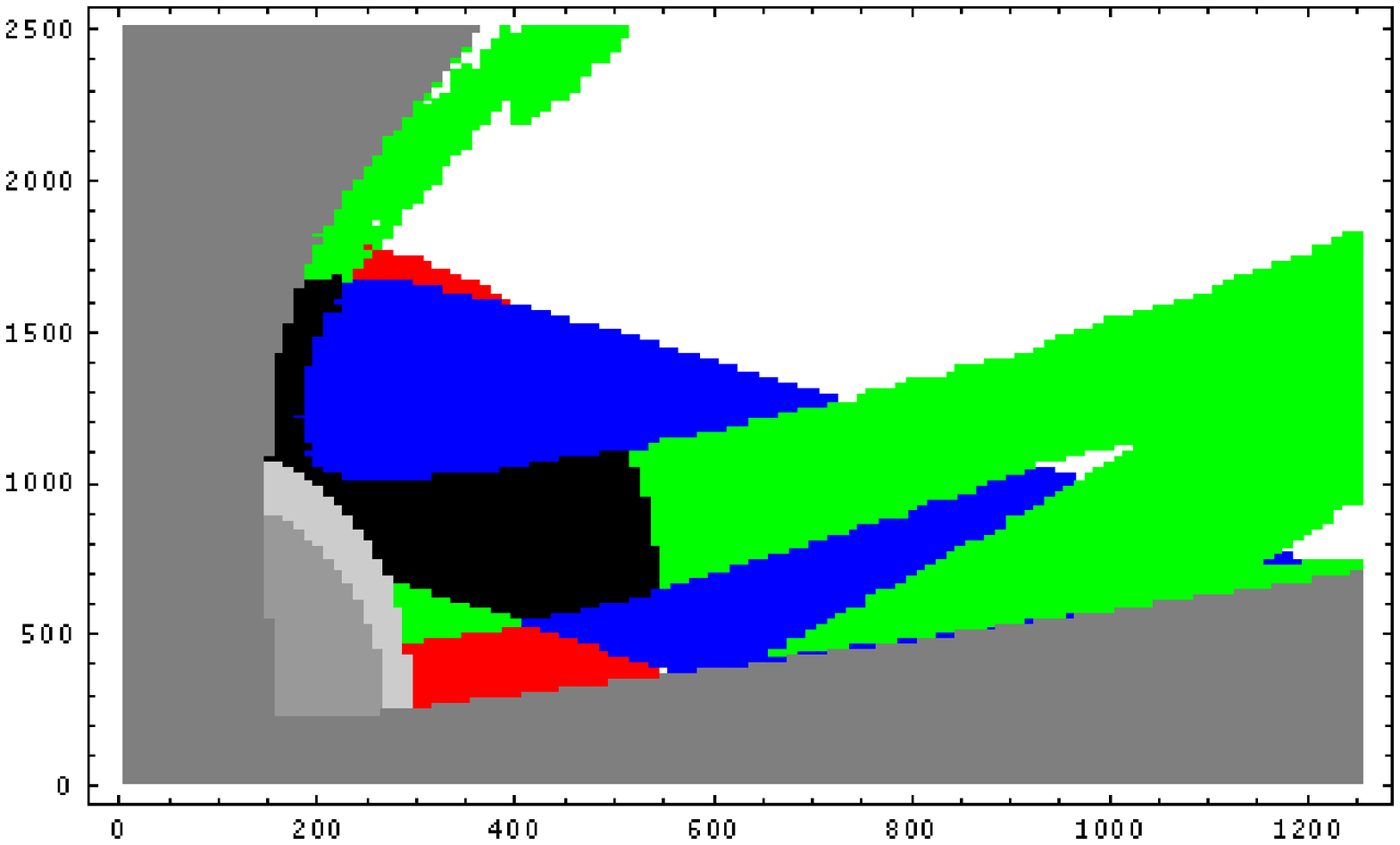,width=0.45\textwidth} 
}
\end{figure}

\vspace*{-5mm} \hspace*{6cm} $m_{1/2}$ \hspace*{7cm} $m_{1/2}$ \\

\vspace*{4mm}
\noindent {\bf Fig.4:} {\it Regions in mSUGRA parameter space allowed
by the DM constraint (green), by the LEP evidence for an SM--like
Higgs boson (red), and by the Brookhaven measurement of $g_\mu -2$
(blue); in the black regions, all three constraints are satisfied
simultaneously, while the grey regions are excluded
experimentally. These figures are for $A_0 = 0, \mu>0$ and $\tan\beta
= 10$ (left) and $60$ (right). }

\vspace*{5mm}

The Dark Matter argument by itself is therefore not sufficient to tell
us in what region of mSUGRA parameter space we are likely to be. The
situation is even worse in other widely discussed, constrained SUSY
models. In models with gauge--mediated SUSY breaking (GMSB), the LSP
is the gravitino $\wt G$, whose mass is essentially independent of the
other sparticle masses.\footnote{Thermal relic gravitinos can form the
Dark Matter if their mass is about 0.5 keV \cite{36}, which implies
that the NLSP is stable at the time scale of accelerator
experiments. Observation of $\lsp \rightarrow \wt G \gamma$ or $\tilde
\tau_1 \rightarrow \wt G \tau$ decays could thus exclude $\wt G$ as DM
candidate.} In anomaly--mediated SUSY breaking (AMSB) the LSP is
wino--like, and makes a good thermal DM candidate for $\mchi \simeq
1.5$ TeV. This prediction is falsifiable in principle, but not at the
any of the colliders that are now under serious consideration.

On the other hand, collider physics experiments can make crucial
contributions to DM physics if \lsp\ indeed constitutes most DM. Once
a DM signal has been established, the first order of business would be
to check whether the WIMP mass is compatible with \mchi\ as measured
at colliders. This crucial check is probably much more demanding for
DM search experiments than for collider experiments. Even at the LHC
it is fairly easy to measure \mchi\ to $\sim 10\%$ \cite{37}. In
contrast, the DAMA ``signal'' \cite{38} only determined the WIMP mass
within a factor of two; the fact that DAMA almost certainly did not
detect WIMPs -- their result is now contradicted by two other
experiments \cite{39}, one of which is background--free -- does not
change the conclusion that determining the WIMP mass from the observed
nuclear recoil spectrum is quite difficult. Fortunately, ``indirect''
WIMP signals, e.g. high--energy neutrinos from WIMP annihilation in
the Sun or Earth, or a $\gamma-$ray line from WIMP annihilation in the
galactic halo \cite{26} -- should more easily allow to determine the
WIMP mass.

Input from collider experiments is also needed for the calculation of
the thermal \lsp\ relic density. In the ``bulk region'' the required
data -- essentially \mchi\ and $m_{\tilde \ell_R}$ -- can already be
provided by the LHC. LHC data can even check whether \lsp\ has
significant higgsino components, which would drastically change the
di--lepton invariant mass distribution, e.g. from $\tilde \chi_i^0
\rightarrow \ell^+ \ell^- \tilde \chi_j^0$ decays \cite{40}. In the
co--annihilation region one would need an accurate measurement of the
$\tilde \tau_1 - \lsp$ mass difference, which is impossible at the LHC
and might even be difficult at future $e^+e^-$ colliders \cite{41}. In
the focus point region one would essentially only need to know the
parameters of the neutralino--chargino system, which are easily
measured at $e^+e^-$ colliders operating at $\sqrt{s} > 2 m_{\tilde
\chi_1^\pm}$ \cite{42}; however, as mentioned above, this may require
a multi--TeV collider if \lsp\ is higgsino--like. Finally, in the
Higgs pole region one would need accurate measurements of the
parameters of both the neutralino and Higgs sectors of the theory. The
latter would be difficult at any planned collider if $m_A$ exceeds 1
TeV or so.

Suppose measurements at colliders indeed allow to predict the thermal
\lsp\ relic density; what would we learn? If the value comes out
higher than the upper bound in (\ref{e9}), we know that the
assumptions going into this calculation are wrong, i.e. either \lsp\
is unstable, or it didn't reach chemical equilibrium with the SM
plasma after the period of last entropy production. Actually, the
latter is not sufficient to avoid ``overclosing the Universe'', since
non--thermal mechanisms can easily over--produce LSPs. This is true in
particular for direct LSP production in the decay of massive particles
whose decays are responsible for the last entropy production,
e.g. inflaton decays \cite{43}. A too high thermal \lsp\ relic density
would thus be evidence (though not proof) that \lsp\ is not stable at
cosmological time scales. This could mean that $R-$parity is not
conserved, or that \lsp\ is actually not the LSP, which instead
resides in the ``hidden sector'' of the theory; this sector interacts
with ordinary (s)particles only through gravitational--strength
interactions. In order to clarify this issue one may then ultimately
have to build an experiment searching for \lsp\ decay with lifetime up
to one minute or so; longer \lsp\ lifetimes are essentially excluded
by combinations of constraints from Big Bang nucleosynthesis and the
spectrum of the cosmic microwave background.

Conversely, if the thermal \lsp\ relic density comes out too low, but
WIMP search experiments indicate that \lsp\ contributes to the DM, one
has proven the existence of non--thermal \lsp\ production
\cite{44,43}, typically from the decay of some more massive
object. Finally, if the thermal \lsp\ relic density indeed comes out
correct, and WIMP search experiments confirm that \lsp\ forms the DM,
we'd have evidence that the Universe reached a re--heat temperature
$T_R \gsim 0.1 \mchi$; note that currently we can only infer \cite{45} 
$T_R \gsim 1$ MeV, from the quantitative success of Big Bang
nucleosynthesis. 

In addition to the thermal relic density, one will also want to
predict the cross section for \lsp--nucleon scattering as well as
$\sigma(\lsp \lsp \rightarrow \gamma \gamma)$. The former determines
the size of both the direct WIMP signal and the signal for WIMP
annihilation in the Earth or Sun (for given WIMP flux), while the
latter is required for the calculation of the size of the most obvious
signal for WIMP annihilation in the halo of the galaxy. If a WIMP
signal has been detected, with kinematics indicating that \lsp\ is
indeed that WIMP\footnote{WIMP discovery by itself should not be
confused with discovery of SUSY. Other WIMP candidates already exist
\cite{46}, and even more models are certain to be constructed as soon
as a signal is found.}, knowledge of these cross section is required
to glean information about the WIMP distribution in the galaxy from
the observed signal(s), which would be of great interest to galactic
modellers; indeed, it might open the door to some sort of ``WIMP
astronomy''. In the absence of a positive signal these cross sections
would indicate the sensitivity an experiment would need to reach in
order to exclude \lsp\ as DM. Unfortunately the \lsp--nucleon
scattering cross section is again sensitive \cite{26} to the Higgs
sector, in particular to the mass and couplings of the heavy neutral
CP--even Higgs boson, which are likely to be among the SUSY parameters
that are most difficult to measure.

As mentioned earlier, most WIMP detection methods can determine the
WIMP mass only with rather large uncertainty. Similarly, the extracted
cross section will come with large uncertainty due to the rather
poorly known \lsp\ flux. As discussed above, the flow of information
regarding these quantities will ultimately be from collider physics to
astro--particle physics, even if the first discovery comes from WIMP
searches. Nevertheless, proof that \lsp\ contributes to the DM would
provide invaluable information to ``mainstream'' particle physics. It
would imply upper limits on the size of all $R-$parity violating
couplings that are many orders of magnitude below the sensitivity of
conceivable laboratory experiments (including proton decay). Moreover,
it would imply that \lsp\ is indeed the LSP, i.e. would impose a
strict lower limit on the mass of any superparticles in the hidden
sector, in particular on $m_{\wt G}$; this information would be of
great interest for both model building and cosmology.

\setcounter{footnote}{0}
\section*{5) The Most Energetic Cosmic Rays}

The most energetic particle collisions that are amenable to
experimental analysis by humans do not occur at colliders; rather,
their origin is the collision of cosmic ray (CR) ``primaries'' with
nuclei in the atmosphere of Earth. The spectrum of these primaries has
now been measured over more than ten orders of magnitude in energy,
and about 30 orders of magnitude in flux. Here we are concerned with
the high--energy end of this spectrum, events with energy around or
above $10^{11}$ GeV. The very existence of such events is puzzling,
for at least two reasons \cite{46a}:

\begin{itemize}

\item It is believed that most CR primaries are accelerated in shock
fronts surrounding celestial bodies. The basic requirement for such
acceleration (or ``bottom--up'') mechanisms to work is that there must
be a sufficiently strong magnetic field $B$ extending over a
sufficient length $L$, i.e. $B \cdot L$ must be sufficiently
large. For example, in a field of $\sim 8$ T extending over $\sim 30$
km, one cannot accelerate protons to more than about 14 TeV; this
describes the LHC, of course. Hillas has shown some time ago \cite{47}
that few, if any, known objects in the Universe have sufficient $B
\cdot L$ to accelerate protons to $10^{11}$ GeV even for perfect
efficiency (i.e. if the $B-$field has collider quality, which does not
strike me as a likely proposition). In such a ``bottom--up'' scenario
neutral primaries would have to be produced when charged particles
collide with ambient matter; these would require even higher energies
for these charged particles than what is observed on Earth.

\item At $E \gsim 5 \cdot 10^{10}$ GeV, protons can photo--produce
pions on the cosmic microwave background, $p + \gamma_{\rm CMB}
\rightarrow \pi + N \ (N = n, \, p)$. A collision of this kind will
reduce the proton's energy by typically 20\% or so. At yet higher
energies, multi--pion production processes, which lead to even higher
energy losses, also become important.\footnote{Protons can also loose
energy through $p + \gamma_{\rm CMB} \rightarrow p + e^+e^-$. This
reaction has much lower energy threshold, but it reduces the proton's
energy only by one percent or so, and is thus of lesser importance.}
The interaction length for such collisions is ``only'' a few tens of
Mpc. As a result, protons with energy above this ``GZK cut--off''
\cite{48} must have been produced within 50 or 100 Mpc of Earth. The
same is true for photons, which can be absorbed by $e^+e^-$ pair
production on background photons in the radio band, as well as heavier
nuclei, which can photo--dissociate on the microwave background, all
with interaction lengths of a few to a few tens of Mpc. On the other
hand, extra--galactic magnetic fields, as well as the fields in our
own galaxy, are so weak that even protons at $10^{11}$ GeV should
still point back to their source (within a few degrees), if it is
within a couple of GZK interaction lengths. However, there are no
nearby potential ``bottom--up'' sources of $10^{11}$ GeV particles
near the measured arrival directions.

\end{itemize}

The solution to these puzzles most likely requires new astrophysics,
new particle physics, or both. There have been several proposals of
how to avoid the GZK cut--off. For example, the primaries might be
\cite{49} ``$R-$hadrons'', containing a light gluino (with mass of a
few GeV). This would push the GZK cut--off to a few times $10^{11}$
GeV, just beyond the energy range covered by current
observations. However, gluinos of the required mass have almost
certainly been excluded by collider experiments
\cite{27,50}. Moreover, this ``solution'' does not address the problem
of how such energies can be reached in the first place. Another
proposal postulates \cite{51} neutrinos as primaries, with mb cross
sections on air. However, models of this kind are severely limited by
unitarity arguments \cite{52}. Besides, the neutrinos would presumably
have to be produced in the collision of even more energetic protons,
thereby {\em aggravating} the first of our two problems. The same is
true for the so--called ``$Z-$burst'' scenario \cite{53}, where one
postulates a very large flux of neutrinos with energies extending to
$10^{12}$ GeV or more impinging on neutrinos in the halo of our
galaxy. These collisions produce on--shell $Z-$bosons, whose decay
into (ultimately) protons and photons (among other things) produces
the CR ``primaries''. It has recently been argued \cite{54,55} that
the required neutrino flux violates current experimental upper limits
(see howeover \cite{55a}). The perhaps most radical
proposal \cite{56} is to simply remove the GZK cut--off by assuming
that Lorentz invariance does not hold at sufficiently high
energies. In this case the photoproduction cross section at $E_\gamma
\sim 10^{-4}$ eV, $E_P \sim 10^{11}$ GeV may not be related to the
measured cross section for $E_\gamma \sim$ GeV on protons at
rest. However, this again does not address the first of our problems.

The to my mind best motivated ``new physics'' explanation is based on
the hypothesis that the highest energy CR events come from the decay
of some very massive $X-$particle. We clearly need $M_X \geq 10^{12}$
GeV, since the most energetic event ever observed \cite{66} has $E
\simeq 3 \cdot 10^{11}$ GeV, and only a (small) fraction of $M_X$ will
go into the energy of any one $X$ decay product. On the other hand,
values of $M_X$ as high as $10^{16}$ GeV are possible. In fact, the
first models of this kind used GUT--scale particles bound in
topological defects \cite{57}; this could ensure their longevity. The
$X-$particles might also roam freely, and be almost stable by virtue
of having extremely suppressed couplings to normal matter
\cite{58}. Several mechanisms have been suggested that allow to
produce sufficiently many such particles at the end of inflation
\cite{59,43}. This solves the first of our two problems: basically
$M_X$ stores energy from an early, extremely violent epoch in the
history of the Universe. The second problem is solved since no
association with known sources of high energy particles is needed
here. Topological defects may not be associated with galaxies. In
contrast, most ``freely'' moving $X-$particles should be enriched in
galaxies \cite{60}, just as other forms of matter are; in fact, they
might even form the Dark Matter. One therefore expects most relevant
decays of freely moving $X-$particles to occur in the halo of our own
galaxy, which predicts an approximately (but not exactly \cite{61})
isotropic distribution of arrival directions, in at least qualitative
agreement with data.

\begin{center} \begin{picture}(-400,300)(0,-190)
\SetPFont{Helvetica}{24}
\PText(-405,4)(0)[]{X}
\DashLine(-410,0)(-440,0){3} \Text(-425,7)[]{$\tilde{q}_L$} 
\DashLine(-400,0)(-370,0){3} \Text(-385,7)[]{$\tilde{q}_L$} 
\Line(-370,1)(-340,26)
\Line(-370,0)(-340,25) \Text(-360,25)[]{$\tilde{g}$} 
\Line(-340,26)(-310,46)
\Line(-340,25)(-310,45) \Text(-330,45)[]{$\tilde{g}$}
\Gluon(-340,25)(-310,10){-3}{4}\Text(-320,25)[]{$g$} 
\DashLine(-310,45)(-280,55){3}  \Text(-295,60)[]{$\tilde{q}_L$} 
\GCirc(-280,55){3}{0}
\Line(-310,45)(-280,35) \Text(-295,30)[]{$q_L$} 
\DashLine(-280,-120)(-280,110){4} \Text(-280,-130)[c]{1 TeV}
\Text(-280,-145)[c]{(SUSY}
\Text(-280,-160)[c]{+ $SU(2)\otimes U(1)$}
\Text(-280,-175)[c] {breaking)}

\Line(-280,55)(-250,80) \Text(-265,80)[]{$q$} 
\Line(-250,80)(-190,70) \Text(-220,68)[]{$q$} 
\Gluon(-250,80)(-220,90){3}{4} \Text(-235,97)[]{$g$} 
\Line(-220,90)(-190,100) \Text(-205,103)[]{$q$} 
\Line(-220,90)(-190,80) \Text(-205,80)[]{$q$} 

\Line(-280,56)(-250,46)
\Line(-280,55)(-250,45)
\Text(-265,42)[]{$\tilde{\chi}_2^0$}
\GCirc(-250,45){3}{0}
\Line(-250,45)(-220,55) \Text(-235,57)[]{$q_L$}
\Line(-250,46)(-170,36)
\Line(-250,45)(-170,35) \Text(-166,35)[l]{$\tilde{\chi}_1^0$}
\Line(-250,45)(-220,25) \Text(-235,27)[]{$q$} 
\Line(-220,55)(-190,65) 
\Gluon(-220,55)(-190,45){-3}{4} 
\DashLine(-190,-120)(-190,110){4} \Text(-190,-130)[c]{1 GeV}
\Text(-190,-145)[c]{(hadronization)}

\Line(-370,0)(-340,-25) \Text(-360,-18)[]{$q_L$} 
\Photon(-340,-25)(-310,-5){3}{5} \Text(-330,-5)[]{$B$} 
\DashLine(-310,-5)(-280,5){3} \Text(-295,8)[]{$\tilde{q}_R$}
\GCirc(-280,5){3}{0}
\DashLine(-310,-5)(-280,-15){3} \Text(-295,-18)[]{$\tilde{q}_R$}
\GCirc(-280,-15){3}{0}
\Line(-340,-25)(-310,-45) \Text(-330,-40)[]{$q$} 
\Photon(-310,-45)(-280,-25){3}{5} \Text(-308,-31)[]{$W$} 
\GCirc(-280,-25){3}{0}
\Line(-280,-25)(-190,-5) \Text(-240,-8)[]{$\tau$} 
\GCirc(-190,-5){3}{0}
\Line(-190,-5)(-160,5) \Text(-175,9)[]{$a_1^{-}$}
\GCirc(-160,5){3}{0}
\Line(-160,5)(-130,15) \Text(-145,17)[]{$\rho^{-}$}
\GCirc(-130,15){3}{0}
\Line(-130,15)(-100,15) \Text(-115,11)[]{$\pi^{-}$}
\GCirc(-100,15){3}{0}
\Line(-100,15)(-70,15) \Text(-66,15)[l]{$\nu_\mu$} 
\Line(-100,15)(-70,5) \Text(-78,2)[]{$\mu^{-}$} 
\GCirc(-70,5){3}{0}
\Line(-70,5)(-50,5) \Text(-46,5)[l]{$\nu_\mu$} 
\Line(-70,5)(-50,-5) \Text(-46,-7)[l]{$\nu_e$} 
\Line(-70,5)(-50,-15) \Text(-46,-16)[l]{$e^{-}$} 

\Line(-130,15)(-100,35) \Text(-115,35)[]{$\pi^0$} 
\GCirc(-100,35){3}{0}
\Photon(-100,35)(-70,35){3}{5} \Text(-66,35)[]{$\gamma$} 
\Photon(-100,35)(-70,55){3}{5} \Text(-66,57)[]{$\gamma$} 

\Line(-160,5)(-130,-5) \Text(-145,-8)[]{$\pi^0$}
\GCirc(-130,-5){3}{0}
\Photon(-130,-5)(-100,-25){3}{5} \Text(-96,-27)[]{$\gamma$} 
\Photon(-130,-5)(-100,-5){3}{5} \Text(-96,-3)[]{$\gamma$} 
\Line(-190,-5)(-160,-15) \Text(-175,-16)[]{$\nu_\tau$} 
\Line(-280,-25)(-100,-45) \Text(-96,-45)[l]{$\nu_\tau$} 

\Line(-310,-45)(-280,-65) \Text(-300,-62)[]{$q$} 
\Line(-280,-65)(-250,-55) \Text(-265,-53)[]{$q$} 
\Gluon(-280,-65)(-250,-75){-3}{4} \Text(-265,-80)[]{$g$}
\Gluon(-250,-75)(-220,-95){-3}{4} \Text(-240,-95)[]{$g$} 
\Line(-220,-95)(-190,-105) \Text(-205,-105)[]{$q$} 
\Line(-220,-95)(-190,-85) \Text(-205,-85)[]{$q$}

\Gluon(-250,-75)(-220,-65){3}{4} \Text(-240,-62)[]{$g$} 
\Line(-220,-65)(-190,-55)  \Text(-205,-55)[]{$q$}
\Line(-220,-65)(-190,-75)  \Text(-205,-75)[]{$q$}

\GOval(-190,65)(20,7)(0){0} 
\GCirc(-170,65){3}{0}
\Text(-175,65)[r]{$n$}
\Line(-170,65)(-130,80) \Text(-126,82)[l]{$p$}
\Line(-170,65)(-130,65) \Text(-126,65)[l]{$e^{-}$}
\Line(-170,65)(-130,50) \Text(-126,48)[l]{$\nu_e$} 

\GOval(-190,-80)(15,7)(0){0} 
\GCirc(-168,-83){3}{0}
\Text(-170,-80)[r]{$\pi^0$}
\Photon(-168,-83)(-130,-65){3}{5} \Text(-126,-63)[]{$\gamma$} 
\Photon(-168,-83)(-130,-95){3}{5} \Text(-126,-98)[]{$\gamma$} 

\end{picture} 
\end{center}
{\bf Fig.~5}: {\it Schematic MSSM ``jet'' for an initial squark with a
virtuality $Q \simeq M_X$. The full circles indicate decays of massive
particles, in distinction to fragmentation vertices. The two vertical
dashed lines separate different epochs of the evolution of the
cascade: at virtuality $Q > m_{\rm SUSY}$, all MSSM particles can be
produced in fragmentation processes. Particles with mass of order
$m_{\rm SUSY}$ decay at the first vertical line. For $m_{\rm SUSY} < Q
< 1$ GeV light QCD degrees of freedom still contribute to the
perturbative evolution of the cascade. At the second vertical line,
all partons hadronize, and unstable hadrons and leptons decay. See the
text for further details.}
\vspace*{5mm}

An $N-$body primary $X-$decay would be like an $N-$jet event produced
at an $e^+e^-$ collider operating at $\sqrt{s} = M_X$. As illustrated
in Fig.~5, such jets would look quite different from the QCD jets we
know. To begin with, the hierarchy between $M_X$ and the weak scale
cries out for supersymmetry as a stabilization mechanism. Since $M_X
\gg$ the sparticle mass scale $m_{\rm SUSY}$, at the early stages of
the jet evolution superparticles can be ``showered off'' just like any
other partons. Moreover, we know that all gauge interactions will have
similar strength at energy or momentum scales near $M_X$. Therefore
electroweak interactions, and perhaps third generation Yukawa
interactions, can play significant roles in the early stages of jet
evolution. However, eventually (after $10^{-27}$ seconds or so) the
shower (virtuality) scale will drop below $m_{\rm SUSY}$. At this
point superparticles will decouple from the shower and decay, possibly
through lengthy decay cascades well known from LHC studies. Slightly
later the same will happen to top quarks as well as electroweak gauge
and Higgs bosons. From this point on electroweak interactions are
indeed far weaker than QCD ones, so that the shower evolution is
dominated by strong interactions. Later yet, all partons will
hadronize. This is the end of jet evolution at collider experiments,
but here we also have to treat the decays of unstable hadrons (mesons,
neutrons etc.) as well as $\tau-$leptons and muons. The final output
of such a calculation \cite{60,62,63} is the spectra of stable
particles ``at source'', i.e. at the location of $X-$decay: electrons,
protons, three kinds of neutrinos (including antiparticles in all
cases), photons, and LSPs. Before comparing with data, propagation
effects may have to be included, especially when one is studying
extra--galactic sources.

One finds that the photon flux at source exceeds the proton flux; this
is in fact also true for jets produced at current colliders. The ratio
is about $1.5:1$ at large $x = E_{\rm particle} / E_{\rm jet}$ if the
jet originates from a (s)quark, about $5:1$ at small $x$ independent
of the particle starting the jet, and exceeds $10:1$ at large $x$ if
the primary particle is a (s)lepton or Higgs boson. This is
problematic, since data indicate \cite{64} that few, if any, of the
highest energy events are due to photons. One thus has to get rid of
most of the photons on the way from the source to Earth. This is
actually expected to occur for sources at cosmological distances, but
would require the galactic radio background to be much stronger than
current best estimates if the sources reside in the halo of our own
galaxy. 

Another practical problem at present is the large discrepancy between
the two measurements with the best statistics of the CR spectrum at
highest energies, by the AGASA and HiRes collaborations
\cite{65}. AGASA sees a spectrum with similar, or perhaps even larger,
spectral index as at lower energy, where the spectrum can be described
by a power law with power $\simeq -2.7$. In contrast, the HiRes
spectrum shows a break at the GZK energy $\simeq 5 \cdot 10^{10}$ GeV,
and drops quickly for $E > 10^{11}$ GeV.\footnote{For some reason the
most recent HiRes spectrum no longer contains the highest energy event
ever observed \cite{66}, by the Fly's Eye collaboration, the direct
predecessor of the HiRes experiment.} The HiRes spectrum can be fitted
\cite{65} assuming a homogeneous distribution of sources with $E^{-2}$
spectrum. However, the fact that such a fit is possible doesn't tell
us much about the actual nature of these sources; the GZK effect
ensures that the information about the spectrum at source with $E
\gsim 10^{11}$ GeV is essentially lost if the sources are distributed
homogeneously. In fact, the HiRes spectrum does not even prove that
sources are distributed uniformly, since a top--down model with $M_X
\simeq 10^{12}$ GeV will also give a rapidly falling spectrum above
$10^{11}$ GeV. Both the HiRes and AGASA spectra can be fitted within
such models. Performing such fits is necessary to fix the overall
normalization of the fluxes of very energetic particles predicted by
these models. However, this normalization is uncertain not only
because of the discrepancy between the (supposedly) best data sets
currently available, but also because the ``background'' is not well
understood. Most people in the field assume that the spectrum below
$10^{10}$ GeV is dominated by conventional (bottom--up) sources, while
top--down models should at least explain the spectrum above $10^{11}$
GeV. However, since we do not know what these ``conventional sources''
actually are, let alone their spectral characteristics, it is not
clear how much of the spectrum between $10^{10}$ and $10^{11}$ GeV
should be explained through $X$ decays. This is of some concern, since
in that energy range the spectrum is known far better experimentally
than at even higher energy.

Nevertheless we can already make semi--quantitative predictions that
should allow to test these models in the not too distant future. In
particular, one expects the neutrino flux to be even higher than the
photon flux, basically because jets contain more charged pions (each
of which produces three neutrinos) than neutral ones (which decay into
two photons each). Neutrino signals at very high energies have
therefore early been recognized as potential test of this kind of
model \cite{67}. More recent calculations \cite{68} indicate healthy
event rates for next--generation experiments like Ice Cube, with
detection possible in the near future in many scenarios. However,
``bottom--up'' models can also lead to significant fluxes of very
energetic neutrinos, either through the interaction of accelerated
protons with material near the source, or through the GZK effect. The
observation of such a neutrino signal may therefore not be sufficient
to discriminate between these classes of models.

On the other hand, only ``top--down'' models predict a significant
flux of LSPs with ultra--high energy. Even though the cm energy of the
collision of a $10^{11}$ GeV proton with a nucleus (or proton) at rest
is much larger than $m_{\rm SUSY}$, only a tiny fraction of such
collisions will produce superparticles, while each collision will
produce (many) pions which decay into neutrinos. Existing bounds on
the flux of very energetic neutrinos thus imply an undetectably small
LSP flux in such models. Even in top--down models one needs an
effective target mass of at least $10^{12}$ tons or so to detect these
very energetic LSPs. The currently best hope for such experiments are
space--based photo--detectors looking for fluorescence light (from the
de--excitation of nitrogen molecules excited in the evolution of showers
in air) in Earth's atmosphere. In order to suppress the much larger
background from neutrinos, one has to look for {\em upgoing}
events. The trick is that (at least for bino--like LSP) there is an
energy range (around $10^9$ GeV) where the Earth is opaque to
neutrinos, but still allows a sizable fraction of the LSP flux to
pass. This should (at least in principle) allow to detect the LSP flux
predicted in many top--down models \cite{69}.

If it could be proven that the most energetic cosmic rays indeed
originate from the decay of very energetic particles, we would have
gained {\em experimental} access to energies far beyond the wildest
dreams of accelerator physicists. This prospect is thrilling indeed,
but we have to keep in mind that this ``access'' is bound to be rather
indirect. Out of several tens of thousands of particles (with $x \geq
10^{-7}$) produced in any given $X$ decay, we at best observe a single
one. In other words, the only measurables are the single--particle
inclusive spectra of protons, neutrinos (alas, with all flavor
information most likely washed out by oscillations), LSPs and,
perhaps, photons (electrons certainly loose their energy through
synchrotron radiation before reaching Earth). These spectra depend on
various aspects of physics at energy scales all the way up to
$M_X$. In particular, we'll want to know the primary $X$ decay
products (in order to ``profile'' the $X-$particle). Another important
question is what kind of interactions are present at energies just
below $M_X$, e.g. whether there are additional gauge interactions
beyond those present in the SM. However, before we can address such
questions we need to have a detailed understanding of physics at scale
$m_{\rm SUSY}$. In particular, sparticle cascade decays will affect
the spectra of neutrinos, LSPs and (to a lesser extent) protons and
photons. In this exciting scenario we would thus need input from
experiments at TeV--scale colliders in order to learn about physics at
energy all the way up to $M_X \gsim 10^{12}$ GeV.

\section*{6) Summary and Conclusions}
In this talk I attempted to describe some of the connections between
astro--particle and collider physics. The observation of neutrino
oscillations gives additional impetus to searches for lepton flavor
violation at accelerators, which in turn can help to pin down
supersymmetric models of leptogenesis. Electroweak baryogenesis in the
MSSM requires a light $\tilde t_1$ squark and sizable CP--violation in
the chargino and neutralino sector, and can thus easily be tested at a
high--energy $e^+e^-$ collider (if not sooner). Information from
collider experiments will be essential to predict the thermal \lsp\
relic density, as well as the cross sections that are needed to
predict the size of WIMP signals; conversely, proof that \lsp\ is
sufficiently long--lived to form the Dark Matter in the Universe will
imply constraints on SUSY parameters that cannot be obtained at
colliders. Finally, if it can be established that the highest energy
cosmic rays originate from the decay of very massive particles,
detailed information about sparticle masses and decay patterns will be
required to deduce the physics at very high energy scales from the
spectra of very energetic protons, neutrinos and LSPs that should
eventually be measured by astro--particle physics experiments. These
kinds of connections should prove fruitful for both fields.

Nevertheless a word of caution might be in order. At present collider
physicists should not put too much stock in ``cosmologically favored''
scenarios or regions of parameter space. All the phenomena discussed
here (neutrino masses, the baryon asymmetry of the Universe, Dark
Matter, and cosmic rays with $E \gsim 10^{11}$ GeV) currently have
several competing explanations, with usually quite different
repercussions for collider physics (or none at all). When planning and
performing experiments at colliders, it is therefore probably better
to simply ignore cosmological considerations, although it might prove
very rewarding to keep these considerations in mind when analyzing the
data. Similarly, astro--particle physicists should be aware that many
limits from collider physics hold only under specific assumptions. To
name an important example, there is {\em no} model--independent limit
on WIMP masses from colliders. However, the interrelations between
astro--particle and collider physics should become very important
indeed once signals replace bounds, i.e. once an unambiguous discovery
of ``new physics'' has been made in either area. Neutrino physics has
shown the way; I fervently hope that additional signals of equal or
even greater importance will be discovered sooner rather than later.

\subsection*{Acknowledgments}
I thank the organizers of the conference for inviting me to give this
talk. I also thank the KIAS school of physics, where I started to
write this up, for their hospitality. Finally, I thank Abdelhak
Djouadi and Jean--Loic Kneur for Fig.~4, and Cyrille Barbot for
Fig.~5. This work is supported in part by the Sonderforschungsbereich
375 Astro--Teilchenphysik der Deutschen Forschungsgemeinschaft.


\begin{thebibliography}{99}
\bibitem{41}
T. Kamon, these proceedings, shows that even identifying $\tilde
\tau_1$ pair production on top of the SUSY background from $\tilde
\chi_2^0 \tilde \chi_1^0$ production will not be easy in such a
scenario.
\end{thebibliography}
\end{document}